# *Ab-initio* Study of Size and Strain Effects on the Electronic Properties of Si Nanowires

X. -H. Peng, [1, 2,*] A. Alizadeh, [3] S. K. Kumar, [4] S. K. Nayak [2]

[1]Department of Applied Sciences and Mathematics, Arizona State University, Mesa, AZ, 85212

[2]Department of Physics, Applied Physics and Astronomy, Rensselaer Polytechnic Institute, Troy, NY, 12180

[3]General Electric Global Research Center, Niskayuna, NY, 12309

[4]Department of Chemical Engineering, Columbia University, New York, NY 10027

* Corresponding author (email): xihong.peng@asu.edu

## ABSTRACT

We have applied density-functional theory (DFT) based calculations to investigate the size and strain effects on the electronic properties, such as band structures, energy gaps, and effective masses of the electron and the hole, in Si nanowires along the <110> direction with diameters up to 5 nm. Under uniaxial strain, we find the band gap varies with strain and this variation is size dependent. For the 1 ~ 2 nm wire, the band gap is a linear function of strain, while for the 2 ~ 4 nm wire the gap variation with strain shows nearly parabolic behavior. This size dependence of the gap variation with strain is explained on the basis of orbital characters of the band edges. In addition we find that the expansive strain increases the effective mass of the hole, while compressive strain increases the effective mass of the electron. The study of size and strain effects on effective masses shows that effective masses of the electron and the hole can be reduced by tuning the diameter of the wire and applying appropriate strain.

# 1. Introduction

The research area of nanoscale semiconductor structures, such as nanowires, has attracted extensive attention over the past several decades[1-11]. Si nanowires, in particular, are expected to play a vital role as both interconnects and functional components in future mesoscopic electronic and optical devices, such as light-emitting diodes (LEDs), field-effect transistors (FETs)[12], inverters[13], and nanoscale sensors[14, 15]. In addition, apart from the possibility of using them in the semiconductor industry, Si nanowires are very interesting for fundamental research, since they provide an opportunity to test quantum mechanical concepts at nanoscale[16].

In recent experiments, researchers were able to make single crystal of Si nanowires with diameters down to 1 nm and lengths of tens of micrometers[4, 17-19]. In these wires, the electric carriers are confined in the lateral direction of the wires, thus quantum confinement effect becomes very important. This effect has been observed, for example, in photoluminescence (PL) studies, and found to exhibit substantial blue-shift of emission with reduction of nanowire diameter[4-7, 17]. For instance, Holmes *et al.*[17] have grown defect-free Si nanowires with nearly uniform diameter (4 nm ~ 5 nm) and length on the order of several micrometers using a supercritical fluid solution-phase approach. They observed visible band-edge PL which was strongly blue-shifted from the bulk Si indirect band gap of 1.1 eV. It was also found that the wavelength of luminescence depends on not only the diameter, but also the crystalline orientation of the wires[5, 6, 17].

In addition to the potential optoelectronic applications, Si nanowires are attractive building blocks for future nano-electronic industry, such as FETs[7, 12, 20] since the reduction in size of a device built from Si nanowires allows increasing speed and computing power, and giving greater device densities. For example, Cui *et al.*[12] reported that Si nanowire FET demonstrates high performance with increase in the average conductance from 45 to 800 nS and average mobility from 30 to 560 $cm^2/V \cdot s$. In addition, these authors also claimed that Si nanowires have the potential to exceed substantially over conventional devices when one compares the transport parameters of scaled Si nanowire FET with those of a state-of-the-art planar metal-oxide-semiconductor FET (MOSFET), and hence could be ideal (best so far) building blocks for future nanoelectronics.

As seen above, much work has been carried out on the electronic properties of nanoscale Si structures. Especially, these studies demonstrated that the electronic properties, such as the energy gap ($E_g$), are strongly dependent on their size. On the other hand, strained Si in the semiconductor industry has been widely used to improve the speed of FETs. For example, in laboratory situations, it has been shown that electrons flow through strained Si 70% faster than in non-strained Si, and strained chip designs can be 35% faster than a standard design, resulting from a lighter effective mass of the electric carrier under strained configuration[21-23]. Si nanowires have attracted much attention due to their potential application in nanoscale circuits and device miniaturization. It is therefore of immense importance to study the strain effects on the effective masses of the electron and the hole, in Si nanowires. In addition, there are several experiments have been carried out to understand the role of strain on the optical emissions of semiconductor nanostructures[24-29]. While these works find evident strain effects on the band gap, there has been no systematic study on the combined effects of size and strain in semiconductor nanowire except for our previous work on Si nanoclusters[30, 31]. Thus it motivated us to systematically study the effects of size and strain on the $E_g$ in Si nanowires. The remaining of our paper is organized as follows. In next section, we present our computational details followed by our results. Section 4 presents our conclusions.

## 2. Simulation details

We have used gradient-corrected (GGA) density-functional theory (DFT)[32] to study the electronic properties of a series of Si <110> nanowires. In particular, we have used Perdew-Wang 91 (PW91) exchange and correlation functional[33] and pseudo-potential plane wave approach with the super cell method. The core electrons are described using ultra-soft Vanderbilt pseudo-potentials[34] within the computational VASP code[35]. The kinetic energy cutoff for the plane wave basis set is 300 eV. The dangling bonds in the wire surface are passivated using hydrogen atoms. Since the wire is one-dimensional nanostructure, the simulation cell along the axial direction is originally taken from Si bulk lattice constant (0.386 nm along the <110> direction), and the lateral size of the cell is chosen so that the distance between the wire and its replica (due to periodic boundary conditions) is more than 0.8 nm. Under this configuration, the interactions between the wire and its replica are negligible. The <110> axial lattice constant is optimized and the total energy of the wire is calculated through energy minimization technique.

The electronic properties of the wire are then calculated. The band gap of the wire is defined by the energy difference between the bottom of the conduction band (conduction band edge – CBE) and the top of valence band (valance band edge – VBE). Once we obtain the band structure of the wire, the effective masses of the electron and the hole can be readily calculated according to the formula $m^* = \hbar^2 \left(\dfrac{d^2\varepsilon}{dk^2}\right)^{-1}$.

Table 1 lists the wires studied in the present work. $N_{Si}$ is the number of Si atoms in a given wire; $N_H$ represents the number of H atoms needed to saturate the surface dangling bonds in the wire; D is the diameter of a wire in the unit of nanometer, and defined as the longest distance between two Si atoms in the wire cross-section; D'(H) is an alternative way to define the diameter of a wire, which measured from the longest distance between two H atoms in the wire cross-section. In this paper, we use the diameter D to define the size of the wire. Fig. 1 gives the snapshots of two Si nanowires at size 1 nm and 2.7 nm viewed from the wire cross-section and side. Blue dots are Si atoms and white are H atoms.

Based on the relaxed wire configurations, we then applied uniaxial strain up to $\pm 3.5\%$ by changing the axial lattice of the wire. The positive values of strain refer to uniaxial expansion, while negative corresponds to compression (note that the lateral *x* and *y* coordinates of the wire are further optimized at a given strain). Our study shows the electronic properties of the wire are affected by the strain. The axial lattice constant, band gap variation with strain, and strain effect on the conduction and valence bands (specifically, the effective masses of the electron and the hole) are reported.

## 3. Results and discussion

### I. Axial lattice constant

We first characterize the structures of the relaxed Si nanowires. The lattice constant $a_{bulk}$ in bulk Si is 0.5461 nm based on the simulation parameters mentioned before. Thus the axial lattice constant $a_{initial}$ of a <110> wire obtained from bulk is 0.386 nm (*i.e.* $a_{initial} = a_{bulk}/\sqrt{2}$). The total energy of the wire is then calculated by relaxing the lateral *x* and *y* coordinates of all atoms. In order to optimize the axial lattice along *z* direction, we performed a series of total energy calculations with different lattice constants. For example, the total energy in the wire of 1.6 nm is plotted as a function of the axial lattice constant in Fig. 2. We find that 0.39 nm is the optimized axial lattice constant instead of the initial 0.386 nm. That means the wire expands along the axial direction upon relaxation. The optimized axial lattice constants for all wires studied in present work are reported in Table 1 (the fifth column). For instance, the optimized lattice constants for the wires of 1.0 nm, 2.2 nm, and 2.7 nm are 0.391 nm, 0.389 nm, and 0.388 nm, respectively. We find that the wires expand along the axial direction compared to bulk Si. In addition, the axial expansion is more apparent for smaller wires and not evident in the wires with diameter larger than 4 nm.

## II. Band structure

Si is an indirect band gap material with the conduction band minima located along the direction Γ to X (*i.e.* π/$a$, where *a* is the lattice constant along axial direction). From the band structure, it has been found that Si <110> wire demonstrates a direct band gap at Γ which is consistent with the literature[5, 6]. For example, in Fig. 3 we show the band structures of the 1.0 nm and 2.7 nm Si wires. It is clear that the conduction band edge (CBE) and valence band edge (VBE) are located at Γ for both of the wires.

Examining the band structures in Fig. 3, we find that the band edges at Γ for the 2.7 nm wire, unlike 1.0 nm nanowire, are degenerate. For example, the CBE is two-fold degenerate. The degeneracy of band edges in a larger wire can be understood as follows. It is known that bulk Si has diamond structure and each Si atom is bonded to other four Si atoms, exhibiting tetrahedral (Td) symmetry. For a Si nanowire, the strict Td symmetry is no longer preserved due to several factors. First, surface Si atoms have dangling bonds. Even when the surface is passivated during the synthesis of the wire, it is difficult to obtain exact sp$^3$ hybridization as in bulk Si. Second, Si wires no longer maintain the same configurations (e.g. bond lengths and bond angles) as bulk Si. For example, in the above subsection of axial lattice constant, the wires expand along axial direction: the expansion is more apparent for smaller wires and negligible for larger wires. However, in a larger nanowire the core Si atoms occupy a significant proportion of the materials, which preserves the symmetry about <110> axis. It has been known that[5, 6] the electron orbitals of the band edges at Γ are mainly contributed by core Si atoms from orbital contour plots of band edge states. Thus the band edges for larger wires are degenerate due to the preserved symmetry.

### III. Band gaps

#### a) Size effects

As mentioned before, the band gap ($E_g$) of a Si wire is defined by the energy difference between CBE and VBE. In Table 1, we report the DFT-GGA predicted band gaps for the Si wires. Note that DFT underestimates band gaps of semiconductors as compared to experimental values of gaps, while more accurate GW method[36-38] and quantum Monte Carlo calculations[5, 39, 40] provide better quantitative predictions. However, previous studies on Si nanoclusters and nanowires showed that the DFT gap predicts a similar size dependency as the optical gap obtained through the GW and quantum Monte Carlo methods[5, 39]. In addition, as reported by Peng *et al*[30], the variation of DFT gap with strain is in excellent agreement with that of the optical gap predicted from these advanced configuration interaction methods. Based on this information we anticipate our DFT results would correctly describe the strain effects on $E_g$ in Si nanowires.

The band gap of Si nanowire in Table 1 is increased when the size of the wire is reduced. This effect is primarily due to quantum confinement. Our predicted size dependence of the band gap in Si nanowires is in a good agreement with the literature[5]. For example, the DFT-LDA predicted energy gaps for the 1.0 nm and 1.6 nm wires in the reference[5] are 1.50 eV and 1.03 eV, and our DFT-GGA predicted gaps are 1.62 eV and 1.18 eV. The slight difference between the studies is due to the different functionals used.

#### b) Strain effects

The results of the effect of strain on the energy gaps in Si wires are presented in Fig. 4. The band gaps as a function of uniaxial strain for several different sized wires are plotted. Negative strain means uniaxial compression and positive strain means uniaxial expansion. For the 1 nm diameter wire, the band gap variation with strain is almost linear. The gap decreases with expansion and increases with compression. The gap variation with strain in the 1.6 nm diameter wire shows a similar linear relation. However, for the 2.7 nm diameter wire, the gap variation with strain shows a nearly parabolic behavior, the gap drops at both compression and expansion. For the intermediate 2.2 nm diameter wire, it behaves *intermediately*, *i.e.* the gap decreases with expansion, while there is nearly no change in $E_g$ with compression. We conclude that the strain effect on the band gap in Si wires is strongly dependent on its size.

In order to understand the size-dependence of the strain effect on the band gap, we examine the variations of the energies of the VBE and CBE with strain. The energies of the VBE and CBE in two wires, whose diameter are 1.0 nm and 2.7 nm, are plotted as a function of strain in Fig. 5. It is clear that the energies of the VBE and CBE in the 1.0 nm diameter wire are almost linear functions of strain. The energies of the VBE and CBE decrease with expansion while increase with compression. In addition, the slope of the CBE plot is slightly smaller (i.e. more negative) than that of the VBE plot. Recall that the energy difference between the CBE and VBE gives the band gap, which is also a nearly linear function of strain (see 1.0 nm graph in Fig. 4). However, for the 2.7 nm wire, the energies of the VBE and CBE are not linear functions with strain. Generally, both energies of the VBE and CBE are reduced under expansion and increased with compression. However, the curve of the CBE decreases faster than that of the VBE under expansion. On the other side, the curve of the CBE increases slower than that of the VBE under compression. A detailed explanation of these different trends is presented later.

In order to explain the trends in Fig. 5, it is necessary to introduce the strain response in the lateral directions (*i.e. x-* and *y*-directions) in the wire when strain is applied to the axial direction (*i.e. z*-direction). As it would be expected, once the axial strain is applied, the bonds in the *x-* and *y*-directions will change and this could be explained by Poisson effect. To quantitatively examine the strain response in the lateral directions, we investigate the structures of nanowires under uniaxial strains. For example, for the 2.7 nm wire, the *x-* and *y*-directions shrink 0.67% and 0.46%, respectively, when 3.5% expansion is applied to the *z*-direction. If this wire is compressed by 3.5% in the *z*-direction, the *x-* and *y*-directions will expand 0.78% and 0.49%. Two general points should be pointed out from the data. First, axial compressive strain causes lateral expansion, while axial expansive strain leads to lateral compression. Second, the responding strains in the lateral directions are smaller than the originating uniaxial strain.

The electron cloud contour plots (*i.e.* iso-value 0.05 surface of the wavefunctions) of the VBE and CBE from the lateral cross-section and side views in the 1.0 nm Si wire are presented in Fig. 6. For both views of the wire, the orbitals of the VBE and CBE have bonding character – the electron cloud is mainly located in the intermediate regions shared by Si atoms. From the above discussion of strain response, the lateral *xy*-plane will bear compressive strain once expansive axial strain is applied to the wire. That means in the *xy*-plane the distance of Si atoms will be reduced. The reduction of Si-Si bond lengths makes the electron cloud of the VBE and CBE orbitals more efficiently shared by Si atoms. This effect results in an increased electron-nucleus Coulomb attraction, thus an appreciable decrease of energies of both the VBE and CBE (the change in the electron-electron repulsion energy is relatively small). In contrast, with uniaxial compression, the lateral *xy*-plane experiences expansive strain. With this expansion, energies of both the VBE and CBE increase due to the decrease of electron-nucleus attraction. This explains the general variation trends of the energies of the VBE and CBE with respect to strain in Fig. 5 – *i.e.* the energies of the VBE and CBE increase with compression while decrease with expansion. In addition, from Fig. 6, we find that the orbital of the CBE is more delocalized than VBE. The orbital of the CBE mainly forms two planar discs and both of them parallel to the lateral *xy*-plane. However, the electron cloud of the VBE is mainly distributed in two oval spheres and a planar disc, where the disc is slightly tilted from the lateral *xy*-plane. Thus, the electron cloud of the CBE is more effectively shared by Si atoms (i.e. more delocalized) in the *xy*-plane compared to that of the VBE. As a result, the energy of the CBE is more sensitive to strain than that of the VBE. Therefore, the slope of the CBE curve in the 1.0 nm wire in Fig. 5 is slightly larger than that of the VBE curve.

For the 2.7 nm wire in Fig. 5, we find the curve of the CBE decreases faster than that of the VBE under expansion, while the curve of the CBE increases slower than that of the VBE under compression. This can be understood from the combined effects of strain and degeneracy of band edges. If we only consider the effect of strain in the larger nanowire, we will expect a similar linear variation of band edges with strain as discussed for the small 1.0 nm wire. However, for the larger wire, the band edges are degenerate due to the symmetry of the core Si atoms. Under uniaxial strain, the symmetry of the core Si atoms is broken and the degeneracy of the band edges is lifted. In this case, the degeneracy lifting of band edges will make the energies of the CBE and VBE vary as parabolic functions of strain [30]. In this parabolic behavior, the energy of the CBE decreases while that of the VBE increases under both expansion and compression (see Reference 30). Thus the curves in Fig. 5 for the larger wire (2.7 nm) can be understood from the combined effects of strain and degeneracy lifting of band edges. A schematic of this combined effect is shown in the Fig. 5(c). For example, under compressive strain, the energy of CBE will reduce due to the degeneracy lifting (parabolic curve), in addition to the increase (linear curve).

These two trends are opposite to each other. Therefore the resulting energy shift of the combined effect becomes smaller at the side of compress (negative) strain. . .

## IV. Effective masses

### a) Size effects

Once we obtain the band structure of a Si wire, we can calculate the effective masses of the electron and the hole according to the definition of effective mass, $m^* = \hbar^2 \left(\frac{d^2\varepsilon}{dk^2}\right)^{-1}$. In detail, we plot the conduction and valence bands as a function of the K vector near $\Gamma$ from -0.1 to +0.1, where ±0.1 is in the unit of $2\pi/a$ ($a$ is the axial lattice constant). Then curves of the energy versus K are fitted using the second order polynomial $\varepsilon = C_1 k^2 + C_2 k + C_3$. From the fitting function, we can obtain the curvature as $C_1 = \frac{1}{2} * \left(\frac{d^2\varepsilon}{dk^2}\right)$. From this, we can further calculate the effective mass of the electron and the hole through the relation $m^* = \hbar^2/2C_1$.

In Table 1, we report the calculated results: $m_e^*$ represents the effective mass of the electron, while $m_h^*$ is the effective mass of the hole. For example, the effective mass of the electron in the 3.3 nm wire is 0.18 $m_e$ and the effective mass of the hole is 0.36 $m_e$. Our predicted effective masses of the electron and the hole in Si <110> wires agree well with earlier results[6]. For example, Vo *et al.*[6] reported that the electron effective mass of Si <110> nanowire of size 1 ~ 3 nm is about 0.12 ~ 0.14 $m_e$, and the hole effective mass is in the range of 0.17 ~ 0.44 $m_e$. In addition, the effective masses of the electron and the hole in Si <110> wires are smaller than those of bulk Si. Note that smaller effective masses of the electron and the hole in a material implies larger electron and hole mobility, and thus increase the speed of devices made from the material. Therefore this finding may be a motivation to use these nanowires as connects or device components in future nanoelectronics, such as FETs.

**b) Strain effects**

We now show the effect of strain on the band structures of Si wires, especially on the valence and conduction bands, and present how strain modifies the effective masses of the electron and the hole. As an example, we will present the strain effects on the conduction and valence bands in the 2.7 nm wire. Note that for other wires similar results are also observed. In Fig. 7, the band energies of the conduction and valence bands are plotted as a function of the K vector from Γ to X under different values of uniaxial strain for the 2.7 nm wire. It is clear that the band energies shift evidently in the near region of Γ, while there are much less shift on other K vectors. Particularly, the energies of the conduction and valance bands at Γ (*i.e.* CBE and VBE) move up under compressive strain, while shift down with tensile strain. Note that these band shifts at Γ under strain have been shown previously in Fig. 5 and they are resulting from the bonding and delocalization characters of the CBE and VBE orbitals. The energy convergence of conduction and valance bands at K vectors away from Γ and its vicinity under strain might be due to the localization character of their orbitals.

Since the uniaxial strain has a dominant effect of shifting energy levels on the bands around Γ (see Fig. 7), it tunes the curvature of the dispersion relation at the band edges. Thus we expect the strain will modify the effective masses of the electron and the hole. In order to calculate the effective masses of the 2.7 nm wire, the dispersion relation at the near region of Γ are plotted under different values of strain (shown in Fig. 8). The effective masses of the electron and the hole are obtained through the parabolic fitting of these dispersion curves. The calculated effective masses of the electron and the hole in the 2.7 nm wire under strain are reported in Table 2. Without uniaxial strain, the effective masses of the electron and the hole are $0.150\,m_e$ and $0.295\,m_e$ respectively. Under 3.5% compressive strain, the effective mass of the electron is increased to $0.632\,m_e$, while the effective mass of the hole is reduced to $0.148\,m_e$. In the contrast, under 3.5% expansive strain, the effective mass of the electron is deceased to $0.138\,m_e$. The effective mass of the hole for the wire under 2.5% and 3.5% strain is not available from the simple effective mass approximation. This point is clearly demonstrated in Fig. 8. Under 2.5% strain, the valence band as a function of the K vector becomes nearly flat and is no longer approximated to a parabolic function near Γ. In this case, the effective mass, $m^* = \hbar^2 \left( \frac{d^2\varepsilon}{dk^2} \right)^{-1}$, becomes a large value.

In order to clearly demonstrate the variation trends of the effective masses of the electron and the hole with strain, we plot the data in Table 2, *i.e.* the effective masses as a function of strain in Fig. 9(c). It shows that the effective mass of the electron (red triangles) increases rapidly with compressive uniaxial strain, while decreases mildly with tensile strain. However, the effective mass of the hole (black dots) reduces under compression, while enhances dramatically with tensile strain. In Fig. 9, we also present the variation of the effective masses of the electron and the hole with strain in three other sized nanowires, namely 1.6 nm, 2.2 nm and 3.3 nm. Similar to the result of 2.7 nm wire, they all show that the expansion increases the effective mass of the hole dramatically while compression reduces it mildly. In addition, compression increases the effective mass of the electron while expansion slightly decreases it.

## 4. Conclusion

In summary, we investigated the size and strain effects on the electronic properties, such as band structures, energy gaps, and effective masses of the electron and the hole, in Si <110> nanowires with diameters up to 5 nm using first principles density functional theory. We find that (1) the nanowires expand along the axial <110> direction compared to bulk Si: the expansion is evident for small wires with diameter less than 4 nm; (2) the band structures of Si <110> wires display direct band gap at Γ; (3) the band gap variation with uniaxial strain is size dependent: for 1 ~ 2 nm wires, the band gap is a linear function of strain while for 2 ~ 4 nm wire, the gap variation with strain shows a nearly parabolic behavior resulting from the localized nature of band edges; (4) strain effects effective masses of the electron and the hole in different manner: expansion increases the effective mass of the hole, while compression increases the effective mass of the electron. Our results of size and strain effects on the band gap suggest that photoluminescence in Si nanowires can be engineered by controlling their size and strain. In addition, the study of size and strain effects on the effective masses of the electron and the hole shows that effective masses of the electron and the hole can be reduced by tuning the diameter of the wire and applying appropriate strain. These results support the motivation for using Si nanowires as components and connects in future nanoelectronics.

## Acknowledgement

This work is supported by the Office of Naval Research (ONR) under Contract No. N0014-06-0481, National Science Foundation (NSF) GOALI award #0327981, IBM Grant No. J71211 and Arizona State University (ASU) Initiative Fund to Peng. We are very thankful to Rensselaer Polytechnic Institute Atlas and ASU Saguaro for providing the computational resources. S. Sreekala, P. Shemella, and F. Tang are acknowledged for helpful discussions.

## Table captions

**Table 1** A list of studied Si nanowires along <110> direction in present work. $N_{Si}$ is the number of Si atoms in a given wire; $N_H$ represents the number of H atoms needed to saturate the surface dangling bonds; D is the diameter of a wire in nanometers measured from the longest distance between two Si atoms in the same layer perpendicular to the wire axis; D'(H) is an alternate way to define the diameter of a wire, which measured from the longest distance between two atoms including both Si and H in the same layer perpendicular to the axis; the fifth column is the optimized axial lattice constants; $E_g$ is the DFT predicted band gaps at Γ; $m_e^*$ and $m_h^*$ are the DFT predicted effective masses of the electron and the hole.

**Table 2** The effective masses of the electron and the hole in the wire with diameter 2.7 nm under different values of uniaxial strain. * means the value is not available through the simple parabolic fitting.

## Figure captions

**Fig. 1** Snapshots of Si nanowires of size 1.0 nm (top) and 2.7 nm (bottom) viewed from the wire cross-section (left) and the side (middle and right, the right column is the snapshots of 6-contiguous simulation cells along the axial *z*-direction). Blue dots are Si atoms, white H atoms. The visualization orientations are also given by the coordinate axes at the left-below corner of each snapshot.

**Fig. 2** The total energy in the <110> wire with diameter 1.6 nm as a function of the axial lattice constant. The axial lattice of 0.390 nm corresponds to the lowest total energy, while the original lattice 0.386 nm obtained from bulk Si corresponds to a higher total energy ($\Delta E_{tol}$ = 0.046 eV).

**Fig. 3** The band structures of two Si wires along the <110> direction: a) 1.0 nm; b) 2.7 nm. Both of them show direct band gaps located at Γ. Note that the band edges at Γ in b) are degenerate due to Td symmetry of the inner Si atoms in the wire.

**Fig. 4** DFT predicted band gap as a function of uniaxial strain for Si wires at different size. Positive strain refers to uniaxial expansion while negative strain corresponds to its compression.

**Fig. 5** The variations of the energies of valence band edge (VBE) and conduction band edge (CBE) in Si nanowires with uniaxial strain: a) 1.0 nm; b) 2.7 nm.

**Fig. 6** Electron cloud contour plots at iso-value 0.05 of the VBE (top) and CBE (bottom) wavefunctions in the 1.0 nm Si nanowire viewed from the lateral *xy*-plane (left) and the side *yz*-plane (right). Red and green colors correspond to positive and negative values of the wavefunctions. Blue dots are Si atoms, white H atoms. The orientation is given by the coordinate axis at the left-below corner of each plot.

**Fig. 7** The energies of the conduction band (left) and valence band (right) are plotted as a function of the K vector from Γ to X under different values of strain for the 2.7 nm Si <110> wire. The uniaxial strain has a dominant effect of shifting energy levels on the conduction and valence bands around Γ.

**Fig. 8** The conduction and the valence bands at the near region of Γ are plotted under different values of strain. The effective masses of the electron and the hole are obtained through parabolic fitting the band edges according to the formula $m^* = \hbar^2 \left(\dfrac{d^2\varepsilon}{dk^2}\right)^{-1}$.

**Fig. 9** The effective masses of the electron and the hole are plotted as a function of uniaxial strain for nanowires at different size, (a) 1.6 nm; (b) 2.2 nm; (c) 2.7 nm; (d) 3.3 nm. It shows that the effective mass of the electron (red triangles) increases rapidly with compressive uniaxial strain, while decreases mildly with tensile strain. However, the effective mass of the hole (black dots) reduces under compression, while enhances dramatically with tensile strain.

| $N_{Si}$ | $N_H$ | D (nm) | D'(H) (nm) | Axial Lattice (nm) | $E_g$ (eV) | $m_e^*$ | $m_h^*$ |
|---|---|---|---|---|---|---|---|
| 16 | 12 | 1.0 | 1.2 | 0.39117 | 1.62 | 0.14 | 0.17 |
| 42 | 20 | 1.6 | 1.8 | 0.39001 | 1.18 | 0.14 | 0.17 |
| 76 | 28 | 2.2 | 2.4 | 0.38924 | 1.02 | 0.14 | 0.26 |
| 110 | 32 | 2.7 | 3.0 | 0.38808 | 0.94 | 0.15 | 0.29 |
| 172 | 44 | 3.3 | 3.6 | 0.38808 | 0.87 | 0.18 | 0.36 |
| 276 | 52 | 4.3 | 4.6 | 0.38615 | 0.80 | | |
| bulk | | | | 0.38615 | 0.65 | 0.24 | 0.46 |

**Table 1, Peng *et al*.**

| | -3.5% | -2.5% | -1.5% | -0.5% | 0.0% | 0.5% | 1.5% | 2.5% | 3.5% |
|---|---|---|---|---|---|---|---|---|---|
| $m_h^*$ | 0.148 | 0.155 | 0.167 | 0.233 | 0.295 | 0.404 | 5.600 | * | * |
| $m_e^*$ | 0.632 | 0.443 | 0.299 | 0.180 | 0.150 | 0.143 | 0.138 | 0.136 | 0.138 |

**Table 2, Peng *et al*.**

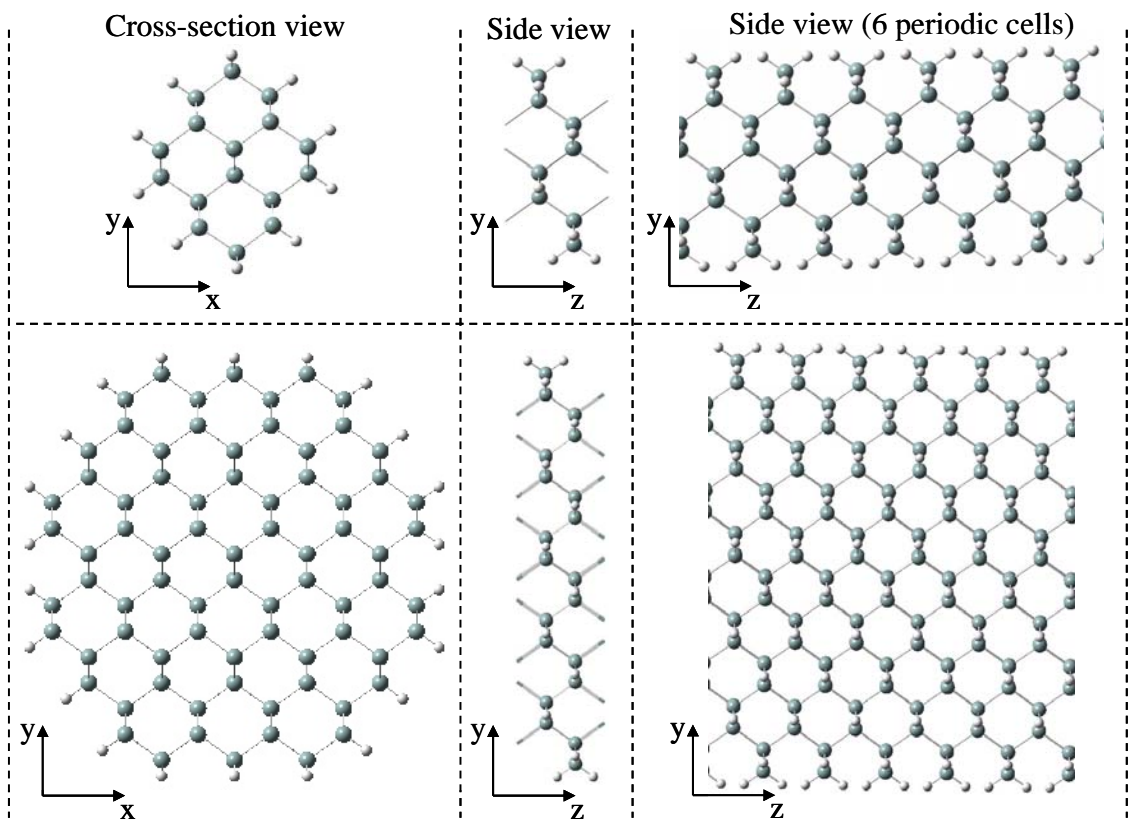

**Fig. 1, Peng *et al*.**

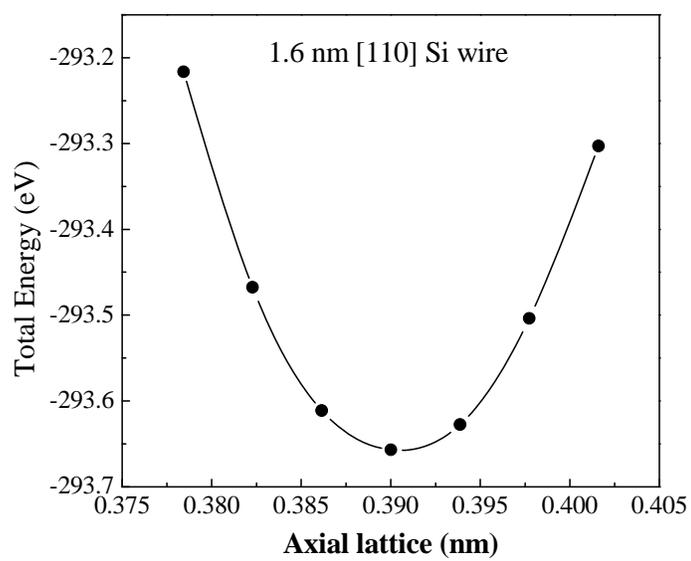

**Fig. 2, Peng *et al*.**

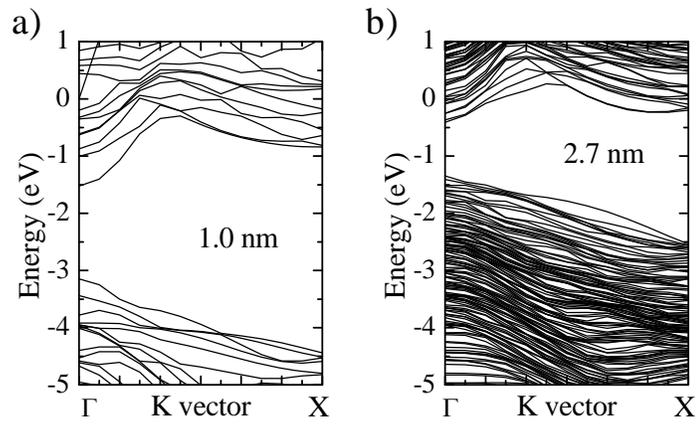

**Fig. 3,** Peng *et al.*

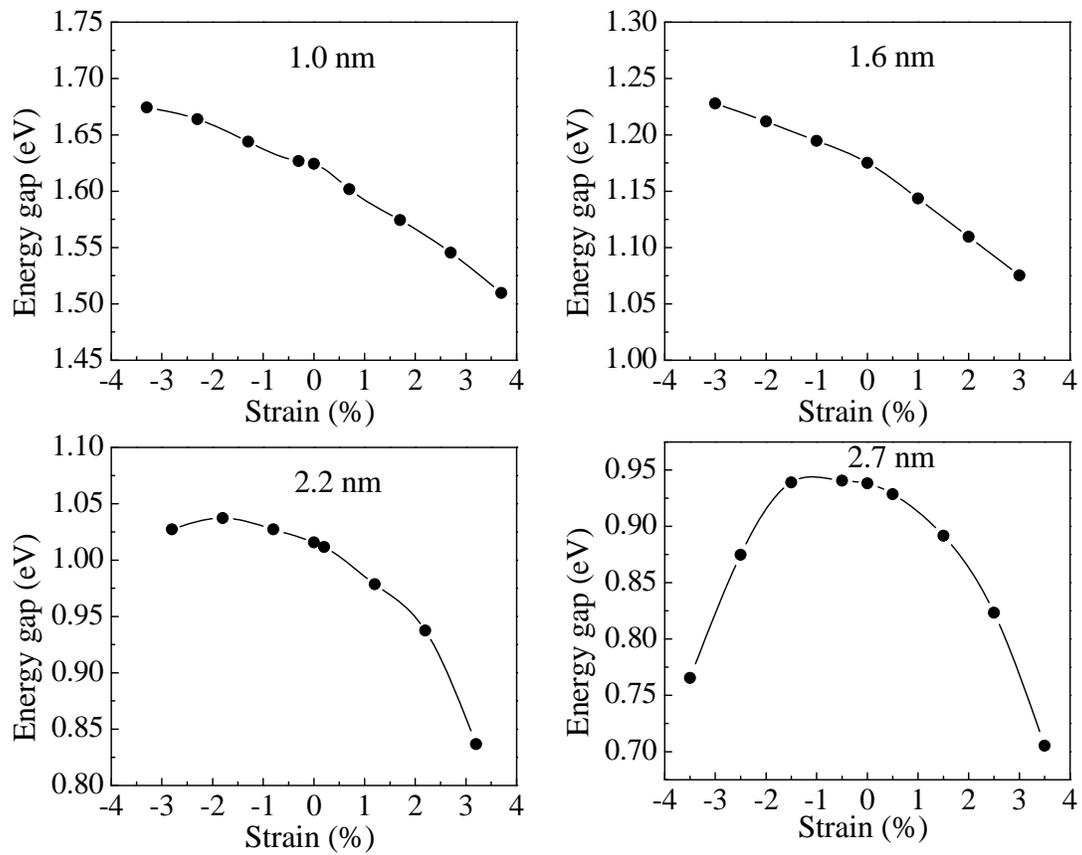

**Fig. 4,** Peng *et al.*

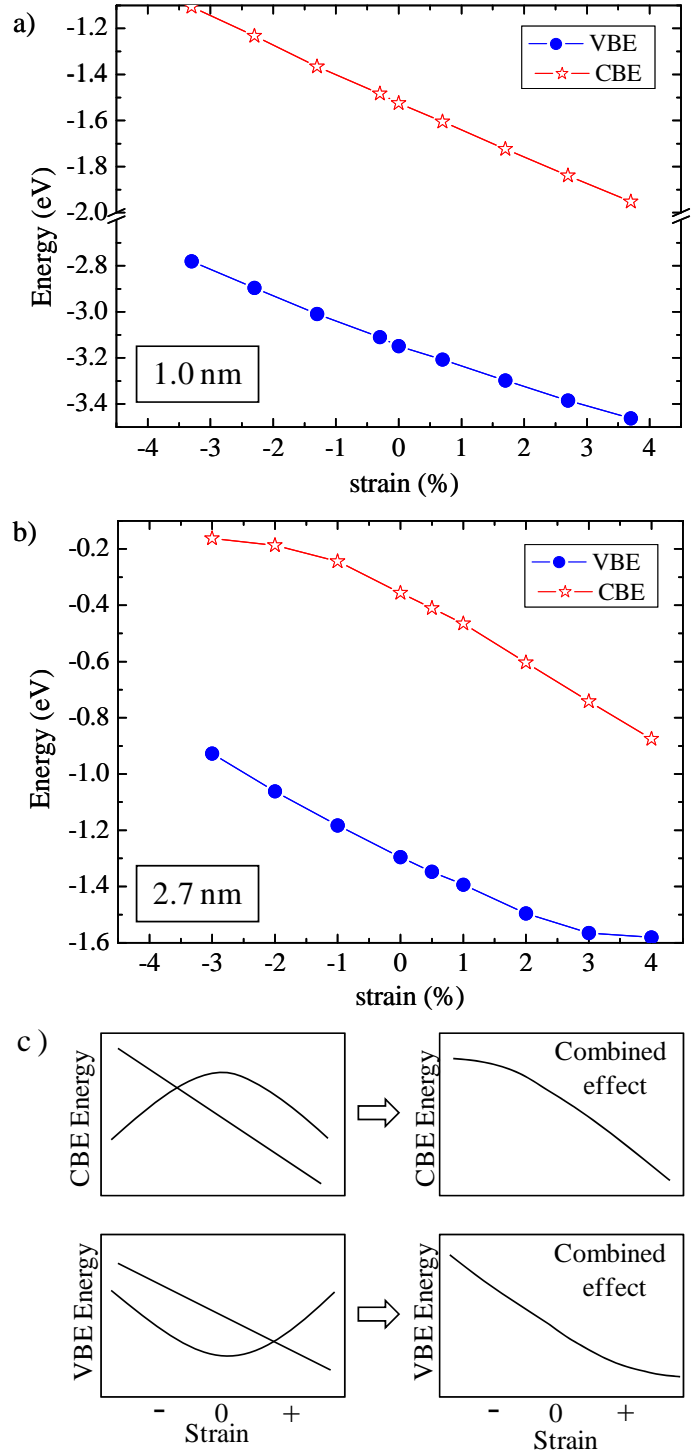

**Fig. 5,** Peng *et al*.

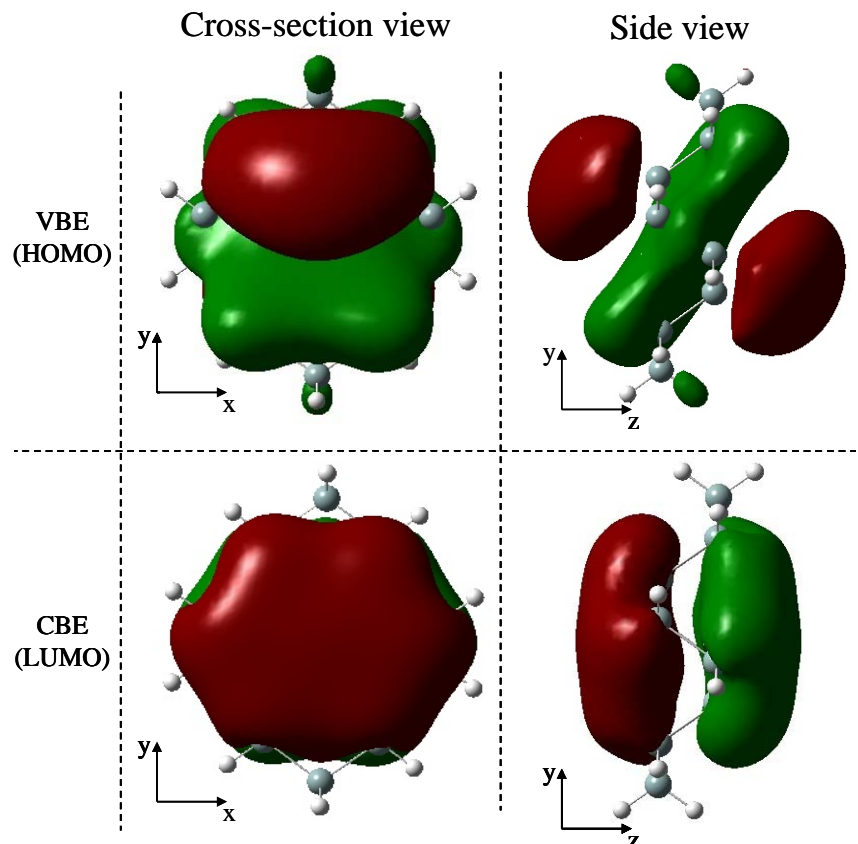

**Fig. 6,** Peng *et al*.

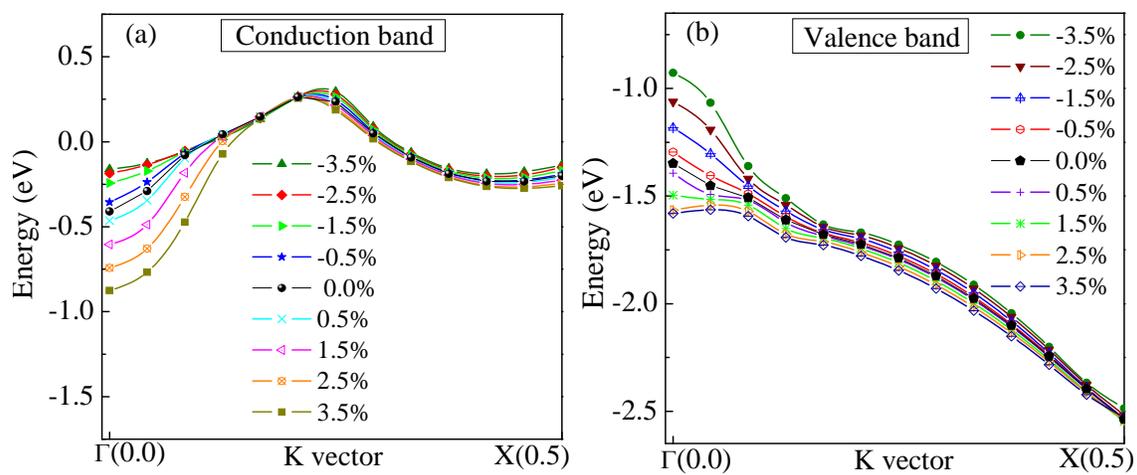

**Fig. 7,** Peng *et al*.

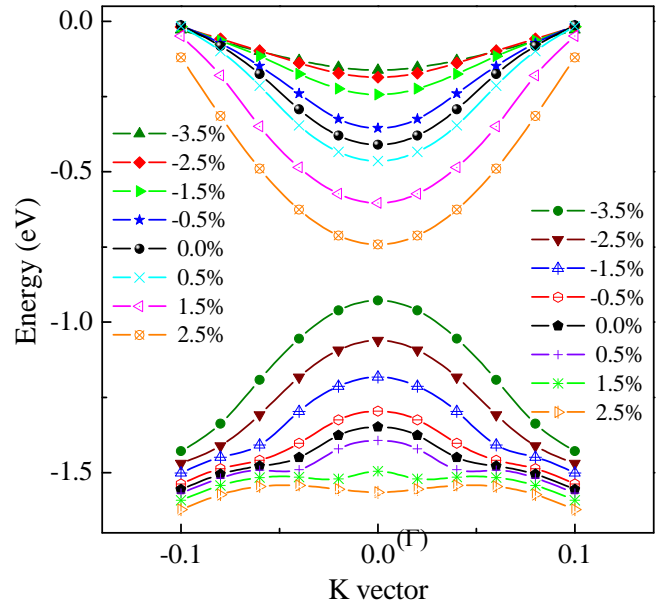

**Fig. 8,** Peng *et al*.

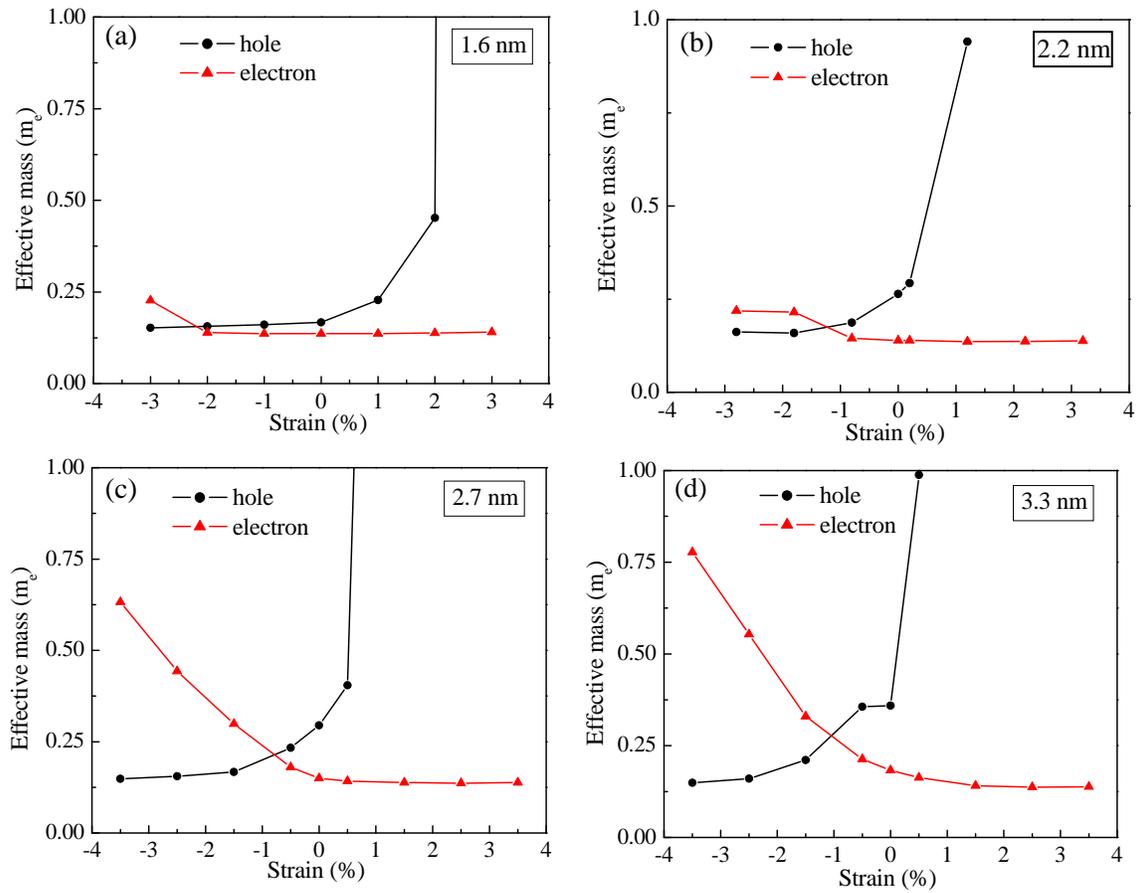

**Fig. 9,** Peng *et al*.